\documentclass[twocolumn,superscriptaddress,pra,aps,showpacs]{revtex4-1}
\usepackage{amsfonts}
\usepackage{amsmath}
\usepackage{amssymb}
\usepackage{graphicx}


\begin{document}

\title{Atomic matter-wave revivals with definite atom number in an optical
lattice}
\author{J.~Schachenmayer}
\affiliation{Institute for Theoretical Physics, University of Innsbruck, A-6020
Innsbruck, Austria}
\affiliation{Institute for Quantum Optics and Quantum Information of the Austrian Academy
of Sciences, A-6020 Innsbruck, Austria}
\author{A.~J.~Daley}
\affiliation{Institute for Theoretical Physics, University of Innsbruck, A-6020
Innsbruck, Austria}
\affiliation{Institute for Quantum Optics and Quantum Information of the Austrian Academy
of Sciences, A-6020 Innsbruck, Austria}
\affiliation{Department of Physics and Astronomy, University of Pittsburgh, Pittsburgh, PA 15260, USA}
\author{P.~Zoller}
\affiliation{Institute for Theoretical Physics, University of Innsbruck, A-6020
Innsbruck, Austria}
\affiliation{Institute for Quantum Optics and Quantum Information of the Austrian Academy
of Sciences, A-6020 Innsbruck, Austria}
\date{\today}

\begin{abstract}
We study the collapse and revival of interference patterns in the momentum
distribution of atoms in optical lattices, using a projection technique to
properly account for the fixed total number of atoms in the system. We
consider the common experimental situation in which weakly interacting
bosons are loaded into a shallow lattice, which is suddenly made deep. The
collapse and revival of peaks in the momentum distribution is then driven by
interactions in a lattice with essentially no tunnelling. The projection
technique allows to us to treat inhomogeneous (trapped) systems exactly in
the case that non-interacting bosons are loaded into the system initially,
and we use time-dependent density matrix renormalization group techniques to
study the system in the case of finite tunnelling in the lattice and finite
initial interactions. For systems of more than a few sites and particles, we
find good agreement with results calculated via a naive approach, in which
the state at each lattice site is described by a coherent state in the
particle occupation number. However, for systems on the order of 10 lattice
sites, we find experimentally measurable discrepancies to the results
predicted by this standard approach.
\end{abstract}

\maketitle

\section{Introduction}

In the early days of Bose Einstein condensates (BECs) with atomic
gases there was a lively debate how to reconcile the notion of the
phase of a BEC with atom number conservation
\cite{Leggett:1991,leggettbook,Javanainen:1996}. In condensed matter
physics this discussion goes back to Leggett's work of spontaneously
broken gauge symmetries, and the observability and meaning of an
absolute phase of a condensate, and of a relative phase between two
BECs \cite{Leggett:1991,leggettbook}. An absolute condensate phase --
provided we understand the BEC order parameter $\langle
\hat{\psi}(\mathbf{x},t)\rangle $ as expectation value of the boson
particle destruction operator -- implies a coherent superposition of
states of atom numbers as in a coherent state.  However, in a theory
where all observables commute with the number operator, the phase of
such superpositions remains unobservable on a fundamental level. On
the other hand, the relative phase of two condensates, associated with
superposition states of atom number differences between two BECs, is
consistent with particle number conservation, and is observable in an
interference experiment which measures a correlation function $\langle
\hat{%
  \psi}^{\dag }(\mathbf{x},t)\hat{\psi}(\mathbf{x}^{\prime },t^{\prime
})\rangle $ \cite{Andrews:1997}. As in the case of coherence of light
in quantum optics \cite{qfcGZ}, the factorization property of the
correlation function $\langle \hat{\psi}%
^{\dag }(\mathbf{x},t)\hat{\psi}(\mathbf{x}^{\prime },t^{\prime
})\rangle \rightarrow \psi ^{\ast }(\mathbf{x},t)\psi
(\mathbf{x}^{\prime },t^{\prime })$ for a given quantum state implies
full visibility of the interference pattern, and provides a
measurement of the BEC order parameter in a number conserving
context. The emergence of interference fringes in a quantum
measurement process based on successive detection of atoms for two
BECs prepared initially in a product of Fock states (eigenstates of
fixed particle number) was analyzed in a language and formalism
familiar from photon detection in quantum optics in
Refs.~\cite{Javanainen:1996}. Detection of an atom in an interference
experiment erases the ``which path'' information from which BEC the
atom came from, and thus prepares a superposition state of atom number
differences compatible with number conservation.

Collapse and revival dynamics due to interactions in a BEC loaded
into an optical lattice, as described by a Hubbard model and observed
in the visibility of interference fringes after releasing atoms from
the trap \cite{Greiner02,Will10}, is a second example where
traditionally atomic number superpositions in the form of coherent
states are invoked in the theoretical derivation and interpretation
\cite{Wright96,Imamoglu97,Milburn97,Walls97}.  Somewhat surprisingly,
a derivation of this phenomenon in a discussion respecting number
conservation seems to be missing in the literature. Below we will
provide analytical collapse and revival formulas for finite particle
number, where we see the emergence of the usual results in the limit
of an infinite number of wells. On the more practical side, we will
also connect the collapse and revival signal with a number conserving
time-dependent DMRG treatment of Bose-Hubbard dynamics in a 1D optical
lattice. This makes it possible to calculate systematically
corrections due to finite tunneling between the wells and finite
initial interactions which has not been possible in previous work.

Collapse and revival of interference patterns driven by two-body
interactions was first raised in discussing the dynamics of Bose-Einstein
Condensates (BECs) in a double well \cite{Milburn97, Walls97}, and later was
investigated with cold atoms in optical lattices \cite{Greiner02, Will10}.
In these lattice experiments, weakly interacting bosons are typically loaded
into a shallow lattice, which is then suddenly made deep. This prevents
tunnelling of atoms between neighbouring sites, and the interference pattern
corresponding to the momentum or quasi-momentum distribution of atoms in the
lattice undergoes collapses and revivals due to on-site interactions. Most
recently, these dynamics were used to investigate the dependence of on-site
interaction energy shifts on the occupation number of a lattice site in a 3D
optical lattice \cite{Will10, Johnson09}. Dynamics of collapses and revivals
driven by two-body interactions have also been studied in the context of
hard-core bosons superlattice \cite{Rigol06}, Bloch oscillations in a tilted
lattice \cite{Kolovsky03, Kolovsky09}, and double well superlattice
experiments, \cite{Anderlini06, Sebby-Strabley07}.

In essentially all of these cases, the experiments have been described based
on an ansatz in which the initial state of the non-interacting BEC can
naively be represented as a product of local coherent states in the onsite
occupation number. This can be represented as 
\begin{align}  \label{eq:naive_ansatz}
|\mathrm{BEC}_{\mathrm{naive}}\rangle = \prod_i^M e^{-\frac{|\beta_i|^2}{2}}
e^{\beta_i b_i^\dag} |\mathrm{vac}\rangle ,
\end{align}
where, $b_i^\dag$ denotes the creation operator for particles on site $i$,
and $|\beta_i|^2$ is the local mean particle number. As we show below, this
ansatz reduces to the correct non-interacting ground state in an infinite
system, but in a finite-size system does not have a fixed total number of
particles.

In a deep lattice, the time-evolution of this state is simply governed
by two-body onsite interactions generated by the Hamiltonian term
$(U/2)\sum_{i}^M \hat n_i(\hat n_i-1)$, with the number operator $\hat
n_i=b_i^\dag b_i$. We find that expectation values such as the
single-particle density matrix (SPDM) factorize, so that for a
homogeneous system, we obtain ($\hbar\equiv 1$):
\begin{align}  \label{eq:naive_spdm_nonhom}
\langle b_{i}^{\dag} b_{j} \rangle_t &= \langle b_i \rangle_t^* \langle
b_j\rangle_t  \notag \\
&= \beta_i^* \beta_j e^{ |\beta_i|^{2}(e^{iUt}-1) } e^{
|\beta_j|^{2}(e^{-iUt}-1) } \\ \label{eq:naive_spdm_hom}
&= |\beta|^2 e^{ |\beta|^{2}(2\cos(Ut)-2) },
\end{align}
with $\beta_i=\beta$ in the homogeneous system.

This naive number non-conserving ansatz leads to wrong results for small
particle numbers, which could be quantitatively measured in an experiment.
In this paper, we demonstrate this by calculating the exact results for
finite systems and a fixed total number of particles via a projection
operator approach. At the same time, we confirm that the naive ansatz yields
accurate results, even for suprisingly moderate-sized systems larger than
about five particles on five lattice sites.

The approach we use here is similar to that used in previous studies
of Bose-Einstein condensates, where a Fock state of fixed particle
number is written as a phase averaged coherent state
\cite{Javanainen:1996}. Our results for a finite system are in
agreement with previous work for homogeneous systems based on a
binomial expansion (see Ref.~\cite{bach2004}), and we are also able to
treat systems with open boundary conditions and trapping
potentials. Using an optimised version of the Time-Evolving Block
Decimation algorithm (TEBD) \cite{Vidal03,Vidal04}, we also study and
discuss the effects of finite tunnelling in the lattice during time
evolution \cite{Fischer08, Wolf10}, as well as weak interactions in
the initial state. Though for clarity we focus on results here in
which the atoms are arranged along one dimension (e.g., with tight
confinement in a 3D optical lattice), the results can be generalized
straightforwardly to higher dimensions.

The rest of this paper is organized as follows: In Sec.~\ref{sec:model} we
first present the details of the system we study, including the Bose-Hubbard
model, and the quasi-momentum distributions and correlation functions we
calculate. In Sec.~\ref{sec:results} we present exact results for revivals
in this system, accounting properly for the conserved total number of
particles in the system, and comparing the results to the naive approach
outlined above. We account also for trapping potentials, finite tunneling in
the lattice, and interactions in the initial state. In Sec.~\ref{sec:summary}
we present a summary and outlook for this work.

\section{The Model}

\label{sec:model} We consider cold atoms loaded into the lowest Bloch band
of an optical lattice \cite{Jak98,Blo08,Lew07}, which are described by the
Bose-Hubbard Hamiltonian ($\hbar=1$) 
\begin{align}  \label{eq:bhham}
H &= 
-J(t) \sum_{\langle i,j \rangle } b_i^{\dag} b_j
+ \frac{U(t)}{2} \sum_{i} \hat n_i(\hat n_i-1) 
+ \sum_{i} \varepsilon_i b_i^\dag b_i,
\end{align}
where $b_i^{(\dag)}$ are the bosonic annihilation (creation) operators at
site $i$, $J$ denotes the tunneling amplitude between neighboring sites, $U$
denotes the onsite interaction energy shift for two atoms, $\hat
n_i=b_i^\dag b_i$ and $\varepsilon_i$ is the energy offset on different
sites, which arises due to external trapping. We will assume that we have a
total of $N$ particles on the $M$ lattice sites.

We assume that at time $t=0$, the system is loaded in the ground state of
the Hamiltonian in a relatively shallow lattice where that atoms are weakly
interacting. At time $t=0$ the depth of the lattice is rapidly increased, on
a timescale that is fast compared with the tunneling time $\sim J^{-1}$, but
slow compared with excitations to higher Bloch bands. In this way, the
parameters $J$ and $U$ are changed suddenly at time $t=0$, so that 
\begin{equation}
J(t)=\left\{ 
\begin{array}{ll}
J_0, & \hbox{$t\leq 0$;} \\ 
J & \hbox{$t>0$.}
\end{array}
\right. , \:\: U(t)=\left\{ 
\begin{array}{ll}
U_0, & \hbox{$t\leq 0$;} \\ 
U & \hbox{$t>0$.}
\end{array}
\right. .
\end{equation}

Below we will first analyze the ideal case, of an initial non-interacting
BEC, so that $U_0=0$, and we have $N$ particles all in the same
single-particle ground state of the Bose-Hubbard model. For a single
particle, the ground state in this system can be written as 
\begin{align}  \label{eq:general_groundstate}
|\varphi\rangle= \frac{1}{\sqrt{N}} \sum_i^M \beta_i \hat b_i^\dag |\mathrm{
vac}\rangle,
\end{align}
with complex amplitudes $\beta_i$ that are dependent on the external
potential $\varepsilon_i$. Note that we normalize these amplitude to the
particle number $N$, i.e.~$\sum_i^M |\beta_i|^2 = N$, for reasons which
become clear below. Then, the initial state for $N$ particles reads 
\begin{align}  \label{eq:initial_BEC_N}
|\mathrm{BEC}_N\rangle = \frac{1}{\sqrt{N! N^N}} \left( \sum_i^M \beta_i
\hat b_i^\dag \right)^N |\mathrm{vac}\rangle.
\end{align}
In the ideal case, after increase of the lattice depth, the tunneling
will be negligible on the timescales of interest after the lattice
depth is increased, i.e., $J\approx 0$. Both of these ideal conditions
will be relaxed in Sec.~\ref{sec:imperfections}.

As mentioned above, collapses and revivals as studied in the experiments are
observed via modulation of the visibility of interference fringes in the
density distribution of the atomic cloud after a free expansion
(time-of-flight measurement). This effectively measures the quasi-momentum
distribution $n_{q}(t)$ after a certain hold time in the lattice $t$,
defined as: 
\begin{align}
n_q(t) \equiv \frac{1}{M} \sum_{i,j}^M \langle b_{i}^{\dag} b_{j}\rangle_t
e^{iq(x_{i}-x_{j})} ,  \label{eq:momentum-distribution}
\end{align}
where $x_i$ are the discrete site-positions in the lattice, and $\langle
b_{i}^{\dagger}b_{j}\rangle_t$ is the time-dependent single particle density
matrix (SPDM). Note that so far we consider a system in arbitrary dimensions
and with arbitrary lattice geometry. Thus, in general the $x_i$ and $q$ are
vectors. Along a single dimension, we consider a lattice constant $a$, and
the discrete positions in that dimension can be written as $x^{\mathrm{1D}
}_i=a i$. Thus, along that dimension, which consists of $M^{\mathrm{1D}}$
sites, the quasi-momenta can have the discrete values $q^{\mathrm{1D}} = n
2\pi / a M^{\mathrm{1D}}$ (with integer $n = 1 \dots M^{\mathrm{1D}}$). In
the following we will focus on the calculation of the time-evolution of the
SPDM elements.

\section{Exact results for Revivals in the Bose-Hubbard model}

\label{sec:reviv-bose-hubb} \label{sec:results}

In this section, we present our results for revivals in the
Bose-Hubbard model, correctly accounting for the fixed total number of
particles in the system. We begin with a simple example of where the
naive treatment breaks down, by studying two particles on two lattice
sites in Sec.~\ref{sec:two}.  In Sec.~\ref{sec:proj} we then provide
technical details of the projection method used to derive exact
results in the case that atoms are initially non-interacting ($U_0=0$)
and that tunneling in the deep lattice is negligible ($J=0$). The
exact results for these cases are then summarized and compared with
the naive approach for homogeneous systems in Sec.~\ref
{sec:homogeneous} and for inhomogeneous systems in Sec.~\ref
{sec:inhomogeneous}. In Sec.~\ref{sec:imperfections} we will relax
these conditions ($U_0=J=0$), accounting for finite tunnelling and
finite interactions in the initial state by performing essentially
exact many-body numerical computations.

\subsection{Two particles on two lattice sites}

\label{sec:two} We begin with a simple example for which the ansatz of
on-site coherent states, Eq.~\eqref{eq:naive_ansatz} clearly gives the
wrong results. This can be easily demonstrated for two particles on
two lattice sites ($N=M=2$) without external trapping, for which the
initial state will be given by $|\mathrm{ \psi_2}\rangle_0=
(|02\rangle + |20\rangle)/2+|11\rangle/\sqrt{2}$, where $ |n_1
n_2\rangle$ denotes a state with $n_1$ particles on the first site and
$n_2$ particles on the second site. Time-evolving this state under the
Hamiltonian \eqref{eq:bhham} gives
\begin{align}
|\mathrm{\psi_2}\rangle_t = \frac{1}{2} \left( e^{-iUt} |02\rangle +
e^{-iUt} |20\rangle \right) + \frac{1}{\sqrt{2}} |11\rangle,
\end{align}
so that the off-diagonal SPDM elements in a two-site system with two
particles are given by 
\begin{align}
\langle b_1^\dag b_2\rangle = \langle b_2^\dag b_1\rangle = \cos{(Ut)}.
\label{eq:two_particles_exact}
\end{align}
This clearly contradicts the result from Eq.~(\ref{eq:naive_spdm_hom}), as
can be seen in Fig.~\ref{fig:spdm_evolution_n}.

\subsection{Particle number projection method}

\label{sec:proj}

In order to conveniently calculate the time evolution for an arbitrary
number of particles, we now introduce an operator $P_{N}$, which
projects states onto a subspace of the Hilbert space with fixed total
particle number $N$. The operator $P_N$ itself does not preserve the
normalization of a state and thus requires an additional
renormalization factor $\mathcal{N}$.  With this operator, we can show
that any state of the form (\ref{eq:initial_BEC_N}) can be written as
a projection of the coherent product state
\begin{align} \label{eq:betai_state} |\{ \beta_k \} \rangle &\equiv
  \prod_{k}^{M} | \beta_k \rangle_k \equiv \prod_{k}^{M} e^{
    -\frac{|\beta_{k}|^{2}}{2} } e^{ \beta_k b_{k}^{\dag} }
  |\mathrm{vac}\rangle .
\end{align}
The number projection of this state reads 
\begin{align}
  P_N |\{ \beta_{k} \} \rangle &= e^{ -\frac{1}{2} \sum_k^M
    |\beta_{k}|^2 }
  P_{N} e^{ \sum_{k}^{M}\beta_k b_{k}^{\dag} }|\mathrm{vac}\rangle  \notag \\
  & = e^{ -\frac{1}{2} \sum_k^M |\beta_{k}|^2 } P_{N}
  \sum_{n=0}^\infty \frac{1}{n!} \left( \sum_{k}^M \beta_k b_k^\dag
  \right)^n |\mathrm{vac}\rangle
  \notag \\
  &= \frac{1}{\sqrt{\mathcal{N}}} e^{ -\frac{1}{2} \sum_k^M
    |\beta_{k}|^2 } \frac{1}{N!} \left( \sum_{k}^M \beta_k b_k^\dag
  \right)^N |\mathrm{vac}  \rangle .
\end{align}
This coincides with the general inital many-body ground state
(\ref{eq:initial_BEC_N}) with the normalization factor
$\mathcal{N}=e^{-N} N^N / N!$, which has the form of a Poisson
distribution in the total particle number $N$. Note that we used
$\sum_k^M |\beta_k|^2 = N$.

To perform calculations conveniently we can write the projection operator in
the form of a phase integral 
\begin{align}  \label{eq:projection_operator}
P_{N} &= \frac{1}{2\pi \sqrt{\mathcal{N}} } \int_{-\pi}^{\pi} d\phi\, e^{ i(\hat{N}-N)\phi } ,
\end{align}
with the total particle number operator $\hat{N} = \sum_{i} b_{i}^{\dagger}
b_{i}$. Note that the projection operator $P_N$ has the properties $%
\mathopen[  P_{N}  , b_{i}^{\dagger} b_{j}  \mathclose] =%
\mathopen[ P_N , H
\mathclose] = 0$ for all $i$, $j$, and $P_{N}^2 = P_{N} / \sqrt{\mathcal{N}}$%
.

The benefit of writing the BEC ground state as number projection of the $|\{
\beta_k \} \rangle$ state, with the projector from Eq.~(\ref%
{eq:projection_operator}) is that the SPDM then factorizes with additional
phase integrals that can be solved exactly. We consider the time-evolved
SPDM under the evolution of Hamiltonian (\ref{eq:bhham}), which is given by 
\begin{align}
\langle b_{i}^{\dag} b_{j}\rangle_t &= \langle\mathrm{BEC}_{N}|e^{iHt}
b_{i}^{\dag} b_{j} e^{-iHt}|\mathrm{BEC}_{N}\rangle  \notag \\
&= \frac{1}{2\pi \mathcal{N} } \int_{-\pi}^{+\pi} d\phi\, e^{-iN\phi}  \notag
\\
& \qquad \times \langle \{ \beta_k \} | e^{iHt} b_{i}^{\dag} b_{j} e^{-iHt}
e^{\hat N \phi} | \{ \beta_k \} \rangle  \notag \\
& = \frac{1}{2\pi \mathcal{N}} \int_{-\pi}^{+\pi} d\phi\, e^{-i N\phi} 
\notag \\
& \qquad \times \langle \{ \beta_k \} | e^{iHt} b_{i}^{\dag }b_{j} e^{-iHt}
| \{ \beta_k e^{i\phi } \} \rangle .  \label{eq:spdm_integral}
\end{align}
In the last line we have used the fact that 
\begin{align}  \label{eq:beta_ephi_trick}
e^{ i \phi b_k^\dag b_k } |\beta_k\rangle_k = \sum_{n=0}^\infty \frac{%
\beta_k^n}{\sqrt{n!}} e^{i \phi n} |n\rangle_k = |\beta_k e^{i \phi}
\rangle_k
\end{align}
for every site $k$.

We can then exactly evaluate the SPDM elements from Eq.~(\ref
{eq:spdm_integral}) for time-evolution under the Hamiltonian
(\ref{eq:bhham}). In the diagonal case, $\mathopen[ H , b_i^\dag b_i
\mathclose] = 0$, and thus the diagonal elements are
time-invariant. At $t=0$, we find
\begin{align}
\langle b_{i}^{\dag} b_{j}\rangle_{t=0} &= \frac{\beta_i^* \beta_j}{2\pi 
\mathcal{N}} \int_{-\pi}^{\pi}d\phi\, e^{ -i(N-1)\phi } e^{
\sum_k^M|\beta_k|^{2}(e^{i\phi}-1) } ,  \label{eq:spdm_t=0}
\end{align}
where we used $b_k|\beta_k\rangle_k = \beta_k |\beta_k\rangle_k$, and $%
\langle \alpha^* | \beta \rangle = e^{\alpha^* \beta - |\alpha|^{2} /2
-|\beta|^{2}/2} $. Integrals of the form of Eq.~(\ref{eq:spdm_t=0}) can be
solved with the parametrization $z=e^{i\phi}$. This yields 
\begin{align}  \label{eq:integral_param}
\langle b_{i}^{\dag} b_{j}\rangle_{t=0} &= \frac{\beta_i^* \beta_j}{2\pi i 
\mathcal{N}} e^{-N} \oint_{\mathcal{C}} dz \, \frac{1}{z^N} e^{N z}  \notag
\\
&= \frac{\beta_i^* \beta_j}{\mathcal{N}} e^{-N} \frac{1}{(N-1)!} \lim_{z \to
0} \left[ \frac{\partial^{N-1}}{\partial z^{N-1}} e^{N z} \right]  \notag \\
&= \beta_i^* \beta_j ,
\end{align}
where the integration takes place along the complex unit circle, denoted by $%
\mathcal{C}$, and we again used $\sum_k |\beta_k|^2 = N$. Thus, the initial
SPDM factorizes into $\beta_i^* \beta_j$, and the diagonal elements remain
constant in time $\langle b_{i}^{\dag} b_{i}\rangle_{t}= |\beta_i|^2$.

We finally calculate the time-evolution of the off-diagonal SPDM elements.
We write the Hamiltonian~(\ref{eq:bhham}) as $H \equiv \sum_i^M h_i $ and
find for $i \neq j$ 
\begin{align}
& \langle \{ \beta_k \} | e^{iHt} b_{i}^{\dag} b_{j} e^{-iHt} | \{ \beta_k
e^{i\phi} \} \rangle  \notag \\
& \quad = \langle \beta_i | e^{ih_{i}t} b_{i}^{\dag} e^{-ih_{i}t} | \beta_i
e^{i\phi} \rangle_{i} \langle \beta_j | e^{ih_{j} t} b_{j}e^{-ih_{j} t} |
\beta_j e^{i\phi}\rangle _{j}  \notag \\
&\qquad \times \prod_{k\neq i,j} \langle \beta_k| \beta_k e^{i\phi}
\rangle_k .
\end{align}

The factorized elements can be straightforwardly calculated as 
\begin{align}
&\langle \beta_j | e^{ih_{j} t} b_{j}e^{-ih_{j} t} | \beta_j
e^{i\phi}\rangle _{j} = \beta_j e^{i\phi} e^{-i\varepsilon_j t} e^{
|\beta_j|^{2} \left[ e^{i\phi} e^{-iUt}-1 \right] } ,
\label{eq:factorized_elements}
\end{align}
and 
\begin{align}
& \langle \beta_i | e^{ih_{i}t} b_{i}^\dag e^{-ih_{i}t} |\beta_i
e^{i\phi}\rangle_{i} = \beta_i^* e^{i\varepsilon_i t} e^{ |\beta_i|^{2} 
\left[ e^{i\phi} e^{iUt}-1 \right] } .  \label{eq:factorized_elements2}
\end{align}
In total we find ($i \neq j$): 
\begin{align}\label{eq:spdm_evolution_full}
\langle b_{i}^{\dag} b_{j}\rangle_t &= \frac{\beta_i^* \beta_j}{2\pi 
\mathcal{N}} \, e^{-N} e^{i (\varepsilon_i - \varepsilon_j) t}
\int_{-\pi}^{+\pi}d\phi \, e^{-i(N-1)\phi}  ) \\
&\quad \times e^{ e^{i\phi} \left [ |\beta_j|^2 e^{-iUt} + |\beta_i|^2
e^{iUt} + \sum_{k \neq i,j} |\beta_k|^2 \right] }  \notag \\
&= \beta_i^* \beta_j e^{i (\varepsilon_i - \varepsilon_j) t}  \notag \\ 
&\hspace{-4ex}\times \left [ \frac{1}{N} \left( |\beta_j|^2 e^{-iUt} + |\beta_i|^2
e^{iUt} + \sum_{k \neq i,j} |\beta_k|^2 \right) \right]^{(N-1)} , \notag
\end{align}
where we again used the parametrization $z=e^{i\phi}$ and an integration
along a complex unit circle.

\subsection{Revivals in a homogeneous system (periodic boundary conditions)}

\label{sec:homogeneous}

In the case of a homogeneous density, $\beta_i=\beta=\sqrt{N/M}$ for
all $i$. This implies that we assume periodic boundary conditions, and
no external trapping ($\varepsilon_i=0$ for all $i$). The off-diagonal
SPDM elements then become site-independent, with
Eq.~(\ref{eq:spdm_evolution_full}) simplifying to
\begin{align}  \label{eq:spdm_evolution_full_hom}
\langle b_{i}^{\dag} b_{j}\rangle_t &= | \beta|^2 \left [ 1 + \frac{|\beta|^2
}{N} \left( 2\cos{(Ut)} - 2 \right) \right]^{(N-1)} .
\end{align}

For $N=M=2$, we reproduce the exact result from Eq.~(\ref
{eq:two_particles_exact}). In the limit where the particle number is
much larger than the onsite density, i.e., $N \gg |\beta|^2$, we find
\begin{align}
|\beta|^2 \left[ 1 + \frac{|\beta|^2}{N} \left( 2 \cos(Ut) - 2 \right) %
\right]^{(N-1)}  \notag \\
\overset{N \gg |\beta|^2}{\longrightarrow} |\beta|^2 e^{ |\beta|^2 \left[ 2
\cos(Ut) - 2 \right] } ,
\end{align}
since $\lim_{N \to \infty} (1+x/N)^N=e^x$. Hence, in this limit we reproduce
exactly the result of the naive treatment from Eq.~(\ref{eq:naive_spdm_hom}).

\begin{figure}[tb]
\includegraphics[width=0.45\textwidth]{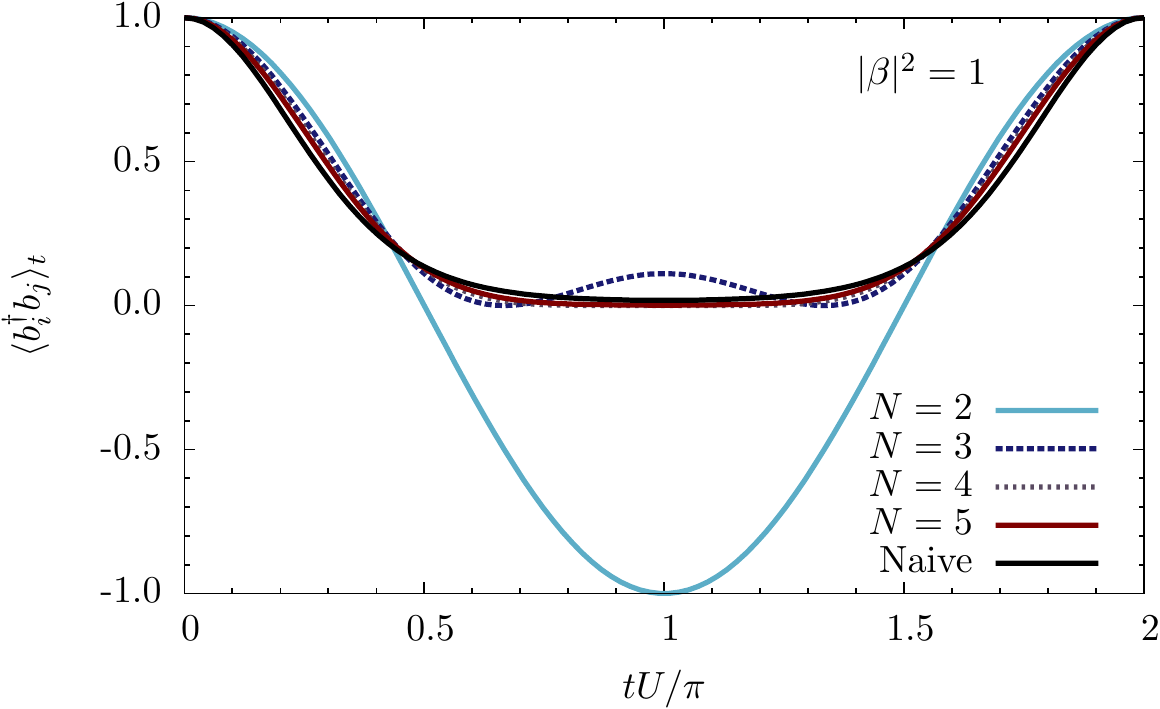}
\caption{Plot of the time-evolution of the off-diagonal SPDM elements
  for an initial ideal superfluid state ($U_0=0$) with homogeneous
  density and without external trapping ($\varepsilon_i=0$) in a deep
  optical lattice ($J=0$). The density is chosen as one particle per
  site ($|\beta|^2=1$), and we show the first collapse and revival
  for increasing number of particles. We compare results from the
  exact treatment of Eq.~(\ref{eq:spdm_evolution_full_hom})
  to the naive result from Eq.~(\ref{eq:naive_spdm_hom})
  (solid black line). Periodic boundary conditions are assumed.}
\label{fig:spdm_evolution_n}
\end{figure}

For homogeneous densities, the height of the $q=0$ peak in the
quasi-momentum distribution is given by $n_{q=0} = (1/M ) \sum_{i,j}^M
\langle b_{i}^{\dag} b_{j}\rangle_t = (M - 1) \langle b_{i}^{\dag}
b_{j} \rangle + |\beta|^2$. Thus, the experimentally measured
visibility of time-of-flight interference fringes is directly
connected to the site-independent off-diagonal SPDM elements. In
Fig.~\ref{fig:spdm_evolution_n} we show a plot of the time-evolution
of the these elements beginning from an ideal non-interacting ground
state with density $|\beta|^2=1$, and including the first collapse and revival. For
increasing particle numbers, we see that the results from
Eq.~(\ref{eq:spdm_evolution_full_hom}) converge rapidly to the values
given by the naive treatment in Eq.~(\ref{eq:naive_spdm_hom}). Indeed,
already for moderate numbers, $N \sim 5$, the exact results are very
close to the approximate result.

We can define the relative error in the visibility of interference fringes,
i.e.~the relative error in the height of the $n_{q=0}$ peak between the
naive and the exact treatment as 
\begin{align}  \label{eq:relerror}
\epsilon_t \equiv \frac{ n_{q=0}^{\mathrm{exact}} - n_{q=0}^{\mathrm{naive}} 
}{ n_{q=0}^{\mathrm{exact}} },
\end{align}
where $n_{q=0}^{\mathrm{exact}}$ is obtained from Eq.~%
\eqref{eq:spdm_evolution_full_hom} and $n_{q=0}^{\mathrm{naive}}$ from Eq.~%
\eqref{eq:naive_spdm_hom}. Using the parametrization $\alpha_t\equiv
|\beta|^2 (2\cos(Ut)-2)$, i.e.~$-4 |\beta|^2 \leq \alpha_t \leq 0$, and
expanding in powers of $1/N$, we find 
\begin{align}  \label{eq:relerror_expand}
\epsilon_t = - \frac{2\alpha_t+ \alpha_t^2}{2 N} - \frac{ 4 \alpha_t^3 +
3\alpha_t^4 }{24 N^2} + \mathcal{O} \left( {\frac{1}{N^3}} \right) .
\end{align}
Thus, for large $N$ and small homogeneous densities, the relative error from
the naive treatment decreases proportional to $1/N$. Note that this error
can also be written in terms of the system size $M$. Up to first order we
find for small densities $|\beta|^2$ 
\begin{align}  \nonumber 
\epsilon_t = - \frac{2\cos(Ut)-2}{M} - \frac{|\beta|^2(2\cos(Ut)-2)^2}{2M} + 
\mathcal{O} \left( {\frac{1}{M^2}} \right) .
\end{align}
The error decreases proportional to $1/M$. In the next section we will also
treat the more general case of inhomogeneous densities below.

\subsection{Revivals for inhomogeneous systems}

\label{sec:inhomogeneous}

Inhomogeneous densities arise from external trapping potentials, i.e.~from
site varying energy-offsets $\varepsilon_i$ in Eq.~(\ref{eq:bhham}), or from
confinement represented by open boundary conditions. The initial state for
non-interacting particles on the lattice can be found by solving the single
particle problem for the amplitudes $\beta_i$. Here we will focus on a 1D
lattice array with $M$ sites, for which the initial single-particle
Hamiltonian reads 
\begin{align} \label{eq:singleham_t<0} H_{t<0}^{} & = -J_0 \sum_{n}^{}
  (|n\rangle\langle n+1| + \mathrm{h.c.} ) + \sum_{n}^{} \epsilon_n
  |n\rangle\langle n| ,
\end{align}
where $|n\rangle$ denotes the state for a particle being at site $n$.
Without external trapping ($\epsilon_n=0$ for all $n$) and for periodic
boundary conditions ($|M+1\rangle\equiv|1\rangle$), the eigenstates of
Hamiltonian (\ref{eq:singleham_t<0}) are simply given by $|\varphi_m\rangle
= \sum_{j=1}^{M} \exp( i q_m a j ) |j\rangle / \sqrt{M}$, with quasi-momenta 
$q_m=2\pi m /aM$ and eigenenergies $E_m=-2\cos(q_ma)$.

For the case of box boundary conditions, we can write the system on a grid
of $M+2$ sites (labeled by $i=0, \dots, M+1$), and local energy offsets $%
\epsilon_0,\epsilon_{M+1} \rightarrow \infty$, and $\epsilon_{n \neq 0,
M+2}=0$. To fulfill the boundary conditions, the eigenstates are then given
by $|\varphi_m\rangle = \sum_{j=0}^{M+1} \sin( q_m a j ) |j\rangle / \sqrt{%
\mathcal{M}}$ with normalization $\mathcal{M} \equiv \sum_{j=0}^{M+1}
\sin^2( q_m a j )$. The quasi-momenta are given by $q_m=m \pi/(M+1)a$, where 
$m>0$, and the eigenenergies are $E_m=-2\cos(q_ma)$. Thus, the coefficients $%
\beta_i$ for the renormalized single-particle ground-state (\ref%
{eq:general_groundstate}) are given by 
\begin{align}  \label{eq:box_boundary_betas}
\beta_i = \sqrt{\frac{N}{\mathcal{M}}} \sin \left( \frac{\pi}{M+1} i \right)
\end{align}

\begin{figure}[tb]
\centering
\includegraphics[width=0.45\textwidth]{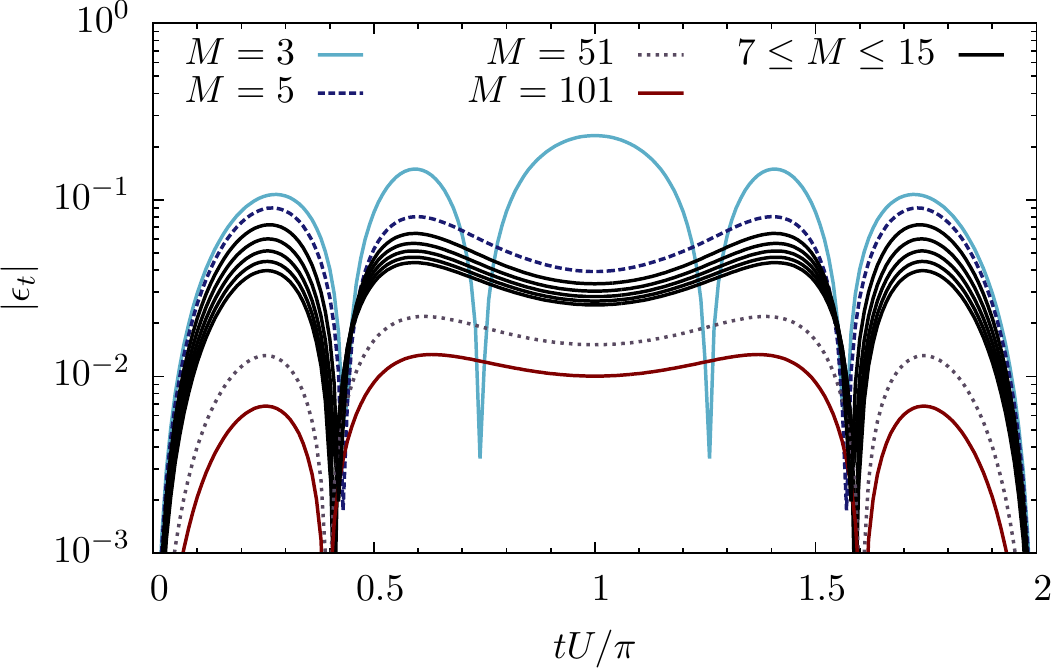}
\caption{The time-evolution of the relative error $\epsilon_t$ of
the naive approach compared with exact results in the height of the
quasi-momentum distribution peak (i.e.~$n_{q=0}$). The results are for a 1D
system with box boundary conditions, resulting in a site-dependent particle
density. Results are shown for the time-evolution of the error until the
first revival at $t=2 \pi/U$, and for increasing system sizes ($N=M$%
, $U_0=0$, $J=0$, $\varepsilon_i=0$).}
\label{fig:nq_evolution_diff}
\end{figure}

In Fig.~\ref{fig:nq_evolution_diff}, we plot the relative error $\epsilon_t$
between the exact and the naive treatment when we begin from an
inhomogeneous initial state due to box boundary conditions. We show the
time-evolution of the error in the height of the $q=0$ quasi-momentum peak,
as defined in Eq.~(\ref{eq:relerror}), covering the time-scale of the first
collapse and revival until $t=2 \pi / U$ in a system with $N=M$ particles.
We find that $\epsilon_t$ generally decreases with increasing system size,
and assumes its largest value approximately in the middle of the collapse
process at $t\approx \pi/4U$.

To analyze the scaling of the error in the particle number $N$, in Fig.~\ref%
{fig:nq_errorscaling}, we plot the relative error in the height of the $q=0$
quasi-momentum peak as a function of $N$. For homogeneous densities in the
limit of large $N$, $\epsilon_{t}$ is given by Eq.~(\ref{eq:relerror_expand}%
), i.e.~it decays proportional to $1/N$. 
\begin{figure}[tb]
\centering
\includegraphics[width=0.45\textwidth]{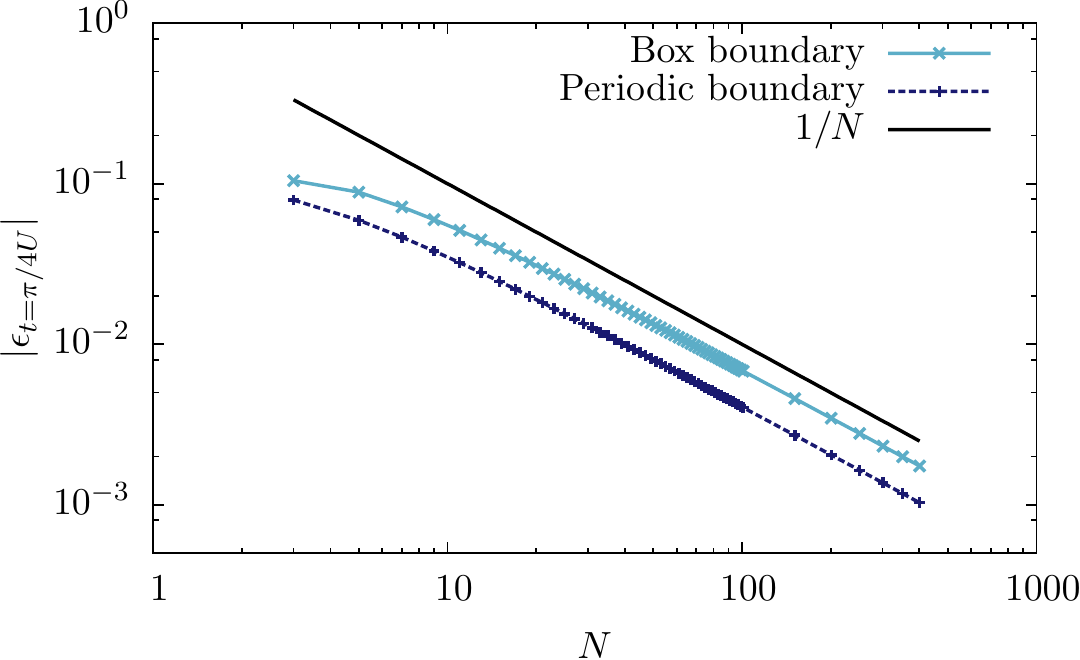}
\caption{The scaling of the relative error $\epsilon_{t=\pi%
    /4U}$ of the naive approach compared with the exact results at the
  middle of the first collapse $t=\pi/4U$. Results are shown both for
  periodic and open (box) boundary conditions. To compare the results
  to Eq.~(\ref{eq:relerror_expand}), we also plot $1/N$ ($U_0=0$,
  $J=0$, $%
  \varepsilon_i=0$).}
\label{fig:nq_errorscaling}
\end{figure}

We thus find that for large $N$, the error scales as $1/N$ for both
homogeneous and inhomogeneous densities studied here. For typical 1D lattice
sizes in experiments of approximately $50$ sites, the relative error becomes
of the order of $1\%$. In this sense, the naive treatment provides a good
semi-quantitative description of the system. However, because experimental
system sizes are of the order of 10--100 lattice sites, it should be
possible to measure the quantitative discrepancies arising in the revivals
as a result of finite system sizes and fixed particle numbers.

\begin{figure}[tb]
\centering
\includegraphics[width=0.4\textwidth]{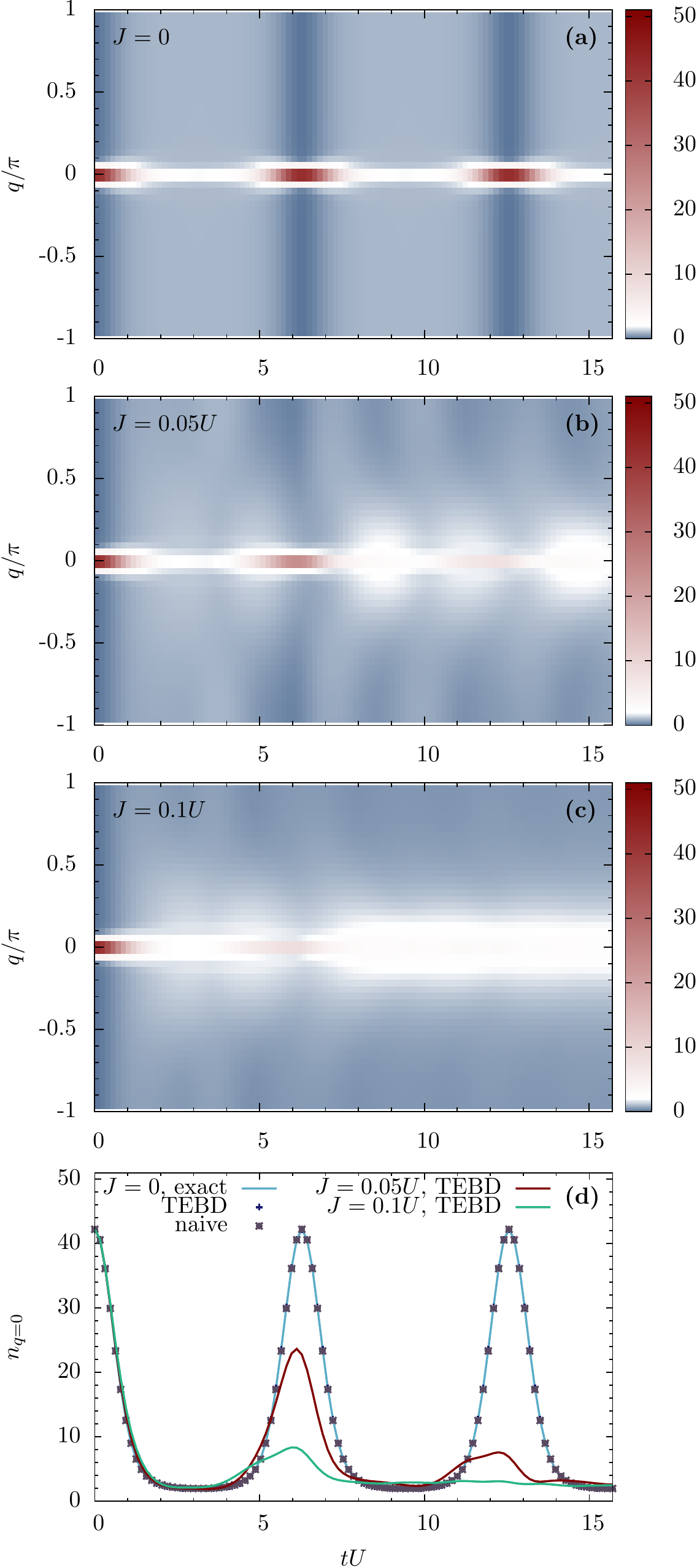}
\caption{Time-evolution of the quasi-momentum distribution $n_q$ for
  several tunneling amplitudes $J$ in a system with $M=51$ sites and
  box boundary conditions. Initially the system is in the exact ground
  state for $U_0/J_0=0$.  Panels (a)-(c) show exact results obtained
  numerically by using the TEBD algorithm. Panel (d) shows the
  evolution of the $n_{q=0}$ peak. The TEBD results for $J=0$ are
  compared to the exact results obtained from
  Eq.~(\ref{eq:spdm_evolution_full}), and to the naive treatment
  obtained from Eq.~(\ref{eq:naive_spdm_nonhom}). The particles are
  intitially non-interacting ($U_0=0$).}
\label{fig:nq_evolution_M51}
\end{figure}

\subsection{Effects of finite tunnelling or initial interactions}

\label{sec:imperfections}

In this section we compare our results to numerical calculations for
the 1D Bose-Hubbard model, in order to account for the effects of a
non-zero tunneling amplitude $J$ during the time-evolution, and a
non-zero initial interaction strength $U_0$. We consider the
time-evolution of the quasi-momentum-distribution in systems up to
$M=51$ sites, using the TEBD algorithm \cite{Vidal03,Vidal04}. This
algorithm makes possible the near-exact integration of the
Schr\"odinger equation for 1D lattice and spin Hamiltonians based on a
matrix product state ansatz. We optimize the algorithm for fixed total
particle numbers as is done in Density Matrix Renormalization Group
methods \cite{dmrgrev,tdmrg1,tdmrg2,tebdcons}

In Fig.~\ref{fig:nq_evolution_M51} we show results for the time evolution of
the quasi-momentum distribution. We start with the exact ground state for
box boundary conditions. It is obtained by an imaginary TEBD time-evolution
and yields an initial density profile, $\beta_i$, consistent with Eq.~(\ref%
{eq:box_boundary_betas}). The exact time-evolution calculation covers the
first two revivals within the time $0 \leq tU \leq 5 \pi \approx 15.7$. In
panels (a)-(c) we show TEBD results for $J=0$, $J=0.05 U$, and $J=0.1 U$,
respectively. In panel (d) we show the time-evolution of the peak of the
quasi-momentum distribution, i.e.~$n_{q=0}$. There, we compare the results
from our TEBD calculations with the exact calculations for $J=0$, i.e.~from
Eq.~(\ref{eq:spdm_evolution_full}), and the naive treatment from Eq.~(\ref%
{eq:naive_spdm_nonhom}). For our numerical calculation, convergence in several
numerical parameters had to be checked. For the results obtained in this
section we have tested convergence of the TEBD algorithm for numerical
parameters, and use matrix sizes up to $\chi=200$ \cite{Vidal03,Vidal04},
and allow occupation numbers up to 14 particles per lattice site.

\begin{figure}[tb]
\centering
\includegraphics[width=0.4\textwidth]{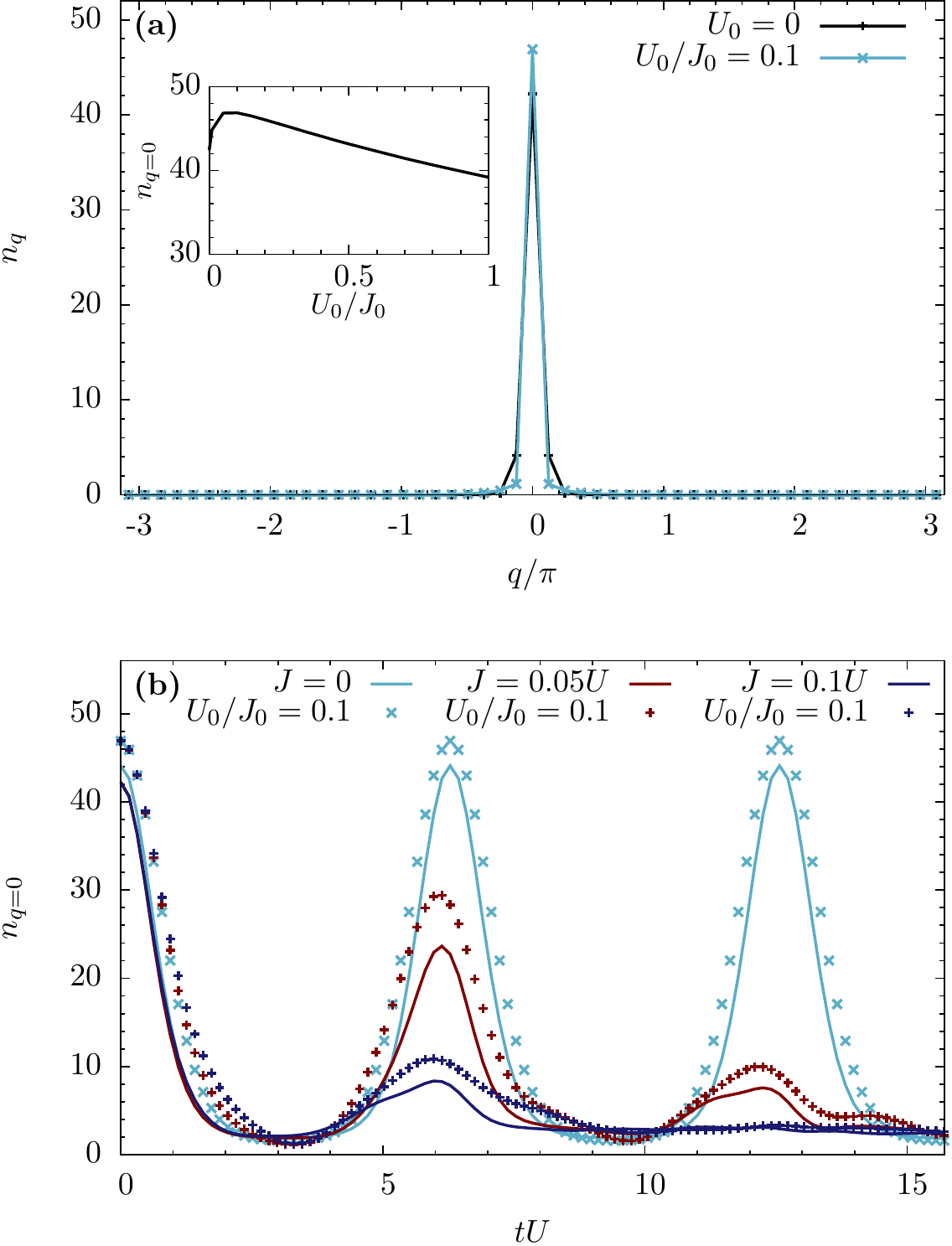}
\caption{Panel (a) compares the initial quasi-momentum distribution for
non-interacting particles ($U_0=0$) with the one for weakly interacting
particles ($U_0/J_0=0.1$). Shown are results for a system with $51$ sites
and box boundary conditions. The inset shows the scaling of the $n_{q=0}$
peak as a function of $U_0$. Panel (b) shows a comparison of the
time-evolution of the $n_{q=0}$ peak for initially non interacting particles
(solid-lines) with weakly interactings ones ($U_0/J_0=0.1$).}
\label{fig:nq_finiteU}
\end{figure}

For $J=0$ results from the analytical number projection method
(Eq.~(\ref{eq:spdm_evolution_full})) perfectly coincide with the TEBD
results as expected, whereas visible differences are observed to the
results from the naive treatment. These errors are small and scale as
in Fig.~\ref{fig:nq_errorscaling}. For non-zero tunneling, we find that the
revivals are strongly damped, i.e.~the peak in the momentum
distribution for $q=0$ is smaller, and the quasi-momentum distribution
at the revival time is broadened compared to the initial one. In
addition, we see that the point in time of the first and the second
revival are markedly shifted towards earlier times for increasing
tunneling strength $J$. Both effects that appear here in exact
numerical calculations of an 1D systems have been recently calculated
perturbatively \cite{Fischer08}, and for hard-core bosons as well as
on a mean-field level in high dimensions \cite{Wolf10}.

For non-zero initial interaction strength $U_0>0$, we note that the
initial quasi-momentum distribution is slightly modified. However, the
basic structure of the revivals is unaffected for $U_0\ll J_0$. This
is shown in Fig.~\ref{fig:nq_finiteU} for a system of $M=51$ sites and
box-boundary conditions. We find that due to boundary effects for $U_0
\lesssim 0.1 J_0$, the quasi-momentum distribution for finite $U_0$
becomes narrowed compared to the one for non-interacting particles
(Fig.~\ref{fig:nq_finiteU}(a)).  Further increasing the initial
interaction strength leads to a depletion of the $q=0$ state and a
broadening of the quasi-momentum distribution (see inset). In
Fig.~\ref{fig:nq_finiteU}(b) we compare the time-evolution for
initially non-interacting particles (solid lines) with the one for $
U_0/J_0=0.1$. We show $n_{q=0}$ for tunneling amplitudes $J/U=0$,
$0.05$, and $0.1$ during time-evolution. We find that the point in
time of the first and the second revival is not altered significantly
due to $U_0 \neq 0$. However, we find that the structure of the peaks
modifies and is smoothed for $J/U>0$.

\section{Summary \& Outlook}

\label{sec:summary}

By using a number projection method, we have shown that exact results can be
derived for the collapse and revival of the SPDM for atoms in an optical
latittce. We properly account for a fixed total number of particles in the
system, and are able to take into account the effects of external
potentials. For large numbers of particles $N$, we find that the discrepancy
to naive calculations, based on assuming a coherent state on each site in
the particle number scales as $1/N$. For as few as 5 particles on 5 sites,
this approach can provide a good semi-quantitative approximation. However,
the discrepancies should be measureable in experiments. Effects of finite
tunnelling in the lattice, which can be treated perturbatively \cite%
{Fischer08}, or exactly for hard-core bosons \cite{Wolf10} can be
investigated directly via time-dependent numerical calculations.

The general technique we use here could be extended to the case in which the
onsite interaction strength is dependent on the occupation number, as has
been observed in recent experiments \cite{Will10}.

\begin{acknowledgments}
  We thank I. Bloch for raising the question of number conservation of
  revivals initially and for discussions.  This work was supported by
  the Austrian Science Foundation through SFB F40 FOQUS, by the
  European union via the integrated project AQUTE, and by the Austrian
  Ministry of Science BMWF via the UniInfrastrukturprogramm of the
  Forschungsplattform Scientific Computing and of the Center for
  Quantum Physics.
\end{acknowledgments}

\end{document}